# Anomalous Spectral Characteristics of Ultrathin sub-nm Colloidal CdSe Nanoplatelets

Sumanta Bose,[1,2,$] Savas Delikanli,[2,3,4] Aydan Yeltik,[4] Manoj Sharma,[4] Onur Erdem,[4] Cuong Dang,[2,3] Weijun Fan,[1,2,*] Dao Hua Zhang,[1,2,3,†] Hilmi Volkan Demir[2,3,4,5,‡]

[1]*OPTIMUS, Centre for OptoElectronics and Biophotonics, Nanyang Technological University, Singapore 639798*
[2]*School of Electrical and Electronic Engineering, Nanyang Technological University, Singapore 639798*
[3]*LUMINOUS! Centre of Excellence for Semiconductor Lighting & Displays and*
*TPI – The Photonics Institute, Nanyang Technological University, Singapore 639798*
[4]*Department of Physics, Department of Electrical and Electronics Engineering and*
*UNAM, Institute of Materials Science and Nanotechnology, Bilkent University, Bilkent, Ankara, Turkey 06800*
[5]*School of Physical and Mathematical Sciences, Nanyang Technological University, Singapore 639798*
E-mail address: [$]*sumanta001@e.ntu.edu.sg*, [*]*ewjfan@ntu.edu.sg*, [†]*edhzhang@ntu.edu.sg*, [‡]*hvdemir@ntu.edu.sg*

**Abstract:** We demonstrate high quantum yield broad photoluminescence emission of ultrathin sub-nanometer CdSe nanoplatelets (two-monolayer). They also exhibit polarization-characterized lateral size dependent anomalous heavy hole and light/split-off hole absorption intensities.
**OCIS codes:** (160.4236) Nanomaterials; (300.0300) Spectroscopy; (350.4238) Nanophotonics and photonic crystals

## 1. Introduction

Quasi-2D atomically-flat colloidal CdSe nanoplatelets (NPLs) have emerged as a new class of novel nanostructures recently [1], with application in high efficiency display devices, LEDs and lasers. The surface effects in ultrathin CdSe NPLs become important as their thickness reduces to sub-nanometer level. Here we study 2-monolayer (2ML) CdSe NPLs of thickness 0.6 nm (Fig. 1a) that exhibit anomalous photoluminescence (PL) and absorption spectra.

## 2. Experimental and Theoretical Results

Colloidal CdSe NPLs are known to have narrow PL linewidths [2], but for 2ML CdSe NPLs, we observed broad photoluminescence emission (Fig. 1b) covering the visible range suitable for white-light emission, with quantum yield (QY) over 85% in solution and 70% in solid film. Such highly efficient broad emission originates from the deep surface trap states as probed by continuous-wave and time-resolved PL (TRPL) techniques. The average carrier lifetime ($\tau_{ave}$) for the pure trap states is 100 ns, while that of the band edge is much shorter, 0.14 ns (Fig. 1c-d) due to carrier trapping. Owing to its low aspect ratio, the number of the trap sites in these ultrathin NPLs is much higher, as a major fraction of the atoms are at the surface. We significantly modified the emission kinetics in 2ML NPLs by using trioctylamine (TOA) as a solvent instead of octadecene (ODE), which enhances the deep trap states QY.

Also, absorption measurements show the anomalous dependence of excitonic transition strength on the NPLs' lateral size (Fig. 1e), which can be controlled by adjusting the growth time after the injection of the Se precursor (TOP-Se) as shown in the TEM images of Fig. 2. The relative strength of the heavy hole (*hh*~393 nm) and light/split-off hole (*lh/so*~372 nm) absorption largely depends on the lateral size of the 2ML CdSe NPLs (Fig. 1f-g), but the spectral position of the peaks stays same throughout. Above a certain lateral size (*ex: Sample* 4), the typical absorption characteristic of thicker NPLs is regained where *hh* peaks are of higher intensity than *lh/so*. To the best of our knowledge, such dependence of excitonic transitions on the lateral size of NPLs has not been reported earlier. Additionally, the 2ML NPLs were observed to exhibit significantly higher intrinsic absorption compared to thicker NPLs (4 and 5 ML) at their *lh/so* absorption peak signifying the presence of giant oscillator strength (GOST) [3].

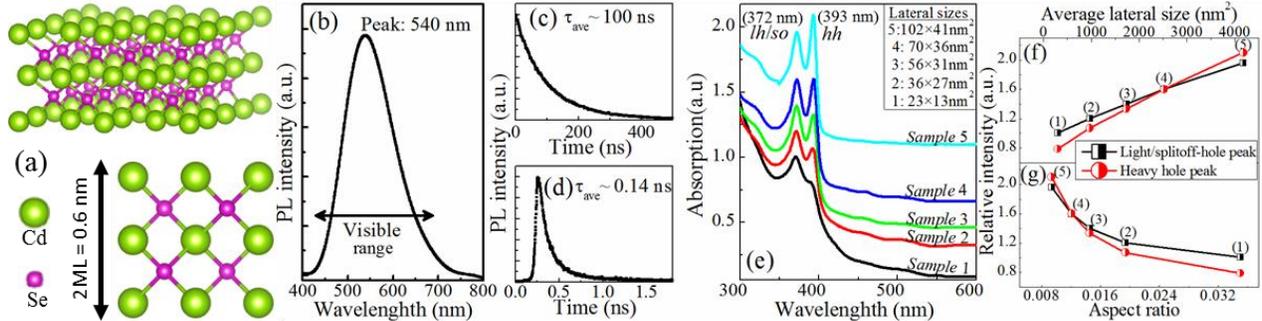

Figure 1: (a) Schematic of a typical 2ML CdSe NPL, (b) Photoluminescence spectrum, (c) TRPL spectrum measured at 550 nm, (d) TRPL spectrum measured at 400 nm, (e) Absorption spectra comparison of 2ML NPLs with varying lateral sizes, and Relative absorption peak intensity of heavy hole (~393 nm) and light/split-off hole (~372 nm) vs. (f) Average NPL lateral area, and (g) NPL aspect ratio=Thickness/√Lateral area.

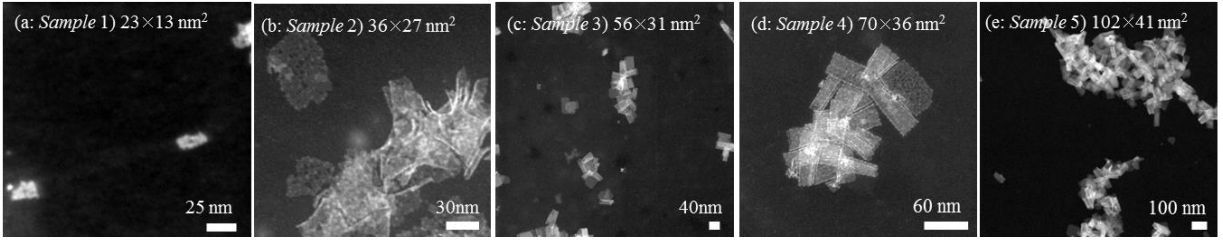

Figure 2: High-resolution Transmission Electron Microscopy (TEM) images of 2ML thick CdSe NPLs with average lateral sizes (a: *Sample* 1) 23×13 nm$^2$, (b: *Sample* 2) 36×27 nm$^2$, (c: *Sample* 3) 56×31 nm$^2$, (d: *Sample* 4) 70×36 nm$^2$, and (e: *Sample* 5) 102×41 nm$^2$. Scales are indicated.

To investigate the origins of such lateral size dependent anomalous characteristics, we studied the five samples of our 2ML CdSe NPLs using an effective mass envelope function theory based on the 8-band $\boldsymbol{k\cdot p}$ model [4]. The 8-band Hamiltonian given by $H = H_k + H_{so} + V_{\text{NPL}}$ simultaneously accounting for the nonparabolicity of the coupled conduction- and valence-band including the orbit-splitting bands was solved, to obtain their electronic bandstructure (*ex*: Fig. 3a for *Sample* 1). The complete absorption spectrum can be calculated as the sum of the bound-state and continuum-state contributions [5], which is strongly dependent on the optical transition matrix element (TME), given by $\mathcal{P}_{cv,i} = \langle \Psi_{c,\mathbf{k}} | e_i \cdot \mathbf{p} | \Psi_{v,\mathbf{k}} \rangle$, $i=x,y,z$ where $\mathbf{p}$ is the momentum operator, $\Psi_{c,\mathbf{k}}$ and $\Psi_{v,\mathbf{k}}$ are the real electron and hole wavefunctions respectively, whose square gives the charge density distribution (i.e. probability of finding *electron* or *hole*). The average of $\mathcal{P}_{cv,x}$ and $\mathcal{P}_{cv,y}$ gives the *x-y* polarized transverse electric (TE) mode TME, while $\mathcal{P}_{cv,z}$ gives the *z* polarized transverse magnetic (TM) mode TME – and they determine the excitonic transition strengths. Numerical results for *Sample* 1 (Fig. 3b-c) show that the TM mode E1-H9 transition TME (=0.69) is stronger than the TE mode E1-H1 transition TME (=0.43). From Fig. 3a, we see the former has greater *lh*/*so* dominance and occurs at 3.27 eV, while the latter has greater *hh* dominance and occurs at 3.1 eV, which is comparable consistent with our experimental observation (Fig. 1e). Both H1 and H9 states have *s*-like spatial charge density pattern (Fig. 3d), overlapping excellently with the E1 state to produce strong excitonic transition, more so for TM E1-H9 than TE E1-H1, as numerically confirmed. Numerical simulations for laterally larger NPLs show rise in both TE and TM mode TME, but the rise in TE mode TME is relatively faster as expected, and exceeds the TM mode TME beyond a certain lateral size – where the typical absorption characteristics become similar to that of thicker NPLs and the *hh* peak becomes larger than the *lh*/*so* peak.

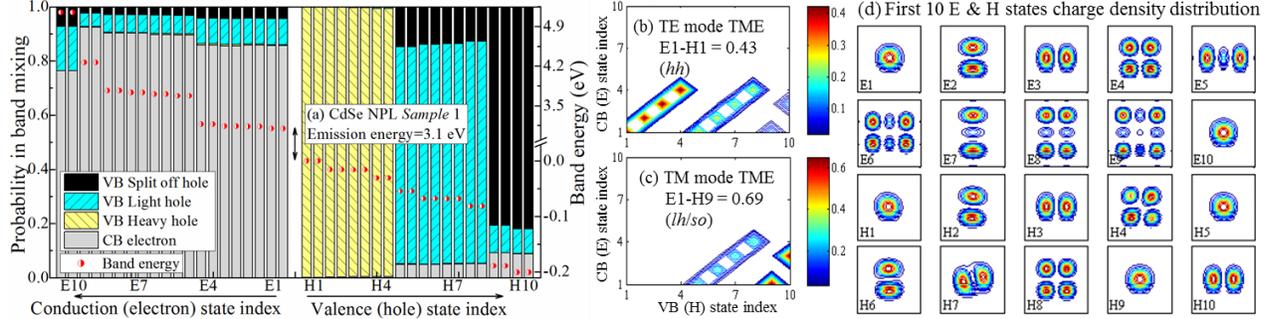

Figure 3: For *Sample* 1, (a) Electronic bandstructure and Probability in band-mixing between conduction electrons and valence *hh*, *lh* and *so* holes showing the first ten conduction (E) states and valence (H) states. Optical transition matrix elements (TME) contour in (b) TE mode i.e. *x-y* polarized, and (c) TM mode i.e. *z* polarized, and (d) Spatial charge density distributions of the first ten conduction (E) and valence (H) states.

### 3. Conclusion and Future works

The relative strength of the *hh* and *lh*/*so* absorption largely depends on the lateral size of the ultrathin sub-nm 2ML CdSe NPLs governed by its TE and TM polarized TMEs. The narrow band-edge absorption suggests that the samples are monodisperse. Therefore, the high QY broad PL emission is the result of a multitude of surface-related trap emitting states. The significantly higher intrinsic absorption, high quantum yield and large Stokes shift make these ultrathin colloidal CdSe NPLs an ideal candidate for optoelectronic applications. Currently, we are making concerted efforts to effectively use TOA as a solvent to improve the QY, and then using them as phosphors integrated as high quality color conversion layers with blue LEDs to produce pleasant and healthy white-light.